\documentclass[12pt, a4paper]{article}
\usepackage{a4wide}
\usepackage{epsf}
\begin{document}
\newcommand{\ua}{|\uparrow\rangle}
\newcommand{\da}{|\downarrow\rangle}

\title{Equal Superposition Transformations \\
 and Quantum Random Walks}
\author{
Preeti Parashar\\
Physics and Applied Mathematics Unit\\
Indian Statistical Institute\\
203 B T Road, Kolkata 700 108, India\\
e-mail: parashar@isical.ac.in
}
\date{}
\maketitle

\begin{abstract}

The largest ensemble of qubits which satisfy the general transformation 
of equal superposition is obtained by different methods, namely, 
linearity, no-superluminal signalling and non-increase of entanglement
under LOCC. We also consider the associated quantum random walk and show
that all unitary balanced coins give the same asymmetric spatial 
probability distribution. It is further illustrated that 
unbalanced coins, upon appropriate superposition, lead to new 
unbiased walks which have no classical analogues.

\end{abstract}

\vskip .8cm


\vskip .5cm


\newpage

\section{Introduction} 

There has been considerable interest in the recent past 
to prove the non-existence of certain quantum unitary operations for 
arbitrary and unknown qubits. Some of the important ones are:
the no-clonning theorem \cite{wz}, the no-deleting principle \cite{pb1},
no-flipping operator \cite{bh} and the no-Hadamard operator \cite{pati1}.
These no-go theorems   
have been re-established by other physical fundamental principles, like 
the no-signalling condition and
no-increase of entanglement under LOCC \cite{gisin}-\cite{preeti}.
 It is then natural to ask that if these operations do not work 
universally (i.e., for all qubits), then for what classes 
of quantum states it would be possible to perform a particular 
task by a single unitary operator. For example, the set of qubits which 
can be flipped exactly by the quantum NOT operator, lie on a great circle 
of the Bloch sphere \cite{pati1, ghosh}. Likewise, the 
largest ensemble of states which can be rotated by the Hadamard gate was 
obtained in \cite{mp}. 

The Hadamard gate creates a superposition of qubit state 
and its orthogonal complement with equal 
amplitudes. In the present work, we consider the most general 
transformation where the    
superposition is with amplitudes which are equal upto a phase.
In other words, the state and its orthogonal superimpose with equal 
probabilities but not necessarily with exactly the same amplitudes.
First, we obtain the largest class of quantum states which can be 
superposed via this transformation. 
Second, it is shown, by using the no-signalling condition and non-increase 
of entanglement under LOCC, that this transformation does not hold for an 
arbitrary qubit. 

The Hadamard transformation is known to be intimately connected to quantum 
random walks which were introduced in \cite{aha}. It has been 
used as a `coin flip' transformation (balanced coin) to study the dynamics 
of such walks \cite{meyer, nayak}. In the same spirit,    
 we consider the quantum random walk associated with our general
transformation and study the probability distribution of the position of a 
particle. It is found that the entire family of such walks gives the 
same asymmetric distribution. We have also considered a unitary 
transformation with unequal amplitudes, serving as an 
unbalanced coin. It is shown that, after a suitable 
superposition, both types of coins lead to symmetric (unbiased) walks. 
However, in the case of unbalanced coin, we obtain new walks that have no 
classical analogues.   

The paper is organized as follows: In Sec. 2 we present the equal 
superposition ensemble. Sec. 3 and 4 pertain to the proving of the 
non-existence of equal superposition transformation for an arbitrary 
qubit. If the state and its orthogonal could be superposed, then it must 
belong to the ensemble presented in Sec. I. This is achieved by imposing
the condition of no-superluminal signalling and non-increase of 
entanglement under LOCC. Sec. 5 is devoted to the study of the associated
quantum random walks. We end the paper with some conclusions in Sec. 6.

\section{The equal superposition ensemble}
  
The computational basis (CB) states $\{ |0\rangle, |1\rangle \}$
of a qubit can be superposed most generally via the  
transformation
\begin{equation}
\label{mgt}
U|0\rangle \rightarrow \alpha |0\rangle + \beta |1\rangle,~~
U|1\rangle \rightarrow \gamma |0\rangle + \delta |1\rangle,
\end{equation}
$\alpha, \beta, \gamma, \delta$ being arbitrary non-zero complex numbers.
We are, however, interested in equal superposition (upto a phase) of the
basis vectors. So let
\begin{equation}
\label{es}
\beta = e^{i \theta} \alpha,~~ \delta = e^{i \phi} \gamma.
\end{equation}
The transformed states are required to be normalized and orthogonal to 
each other. This imposes the following constraints
\begin{equation}
\label{esc}
\alpha \alpha^* = \gamma \gamma^* = 1/2,~~~
\phi = \theta + \pi.
\end{equation}
Eq.(\ref{mgt}) then becomes 
\begin{equation}
\label{est}
U|0\rangle \rightarrow \alpha |0\rangle + e^{i \theta} \alpha |1\rangle,~~
U|1\rangle \rightarrow \gamma |0\rangle - e^{i \theta} \gamma |1\rangle,
\end{equation}
with the unitary matrix given by 
\begin{equation}
\label{esm}
U= \left[\begin{array}{crr}
\alpha & \gamma \\
e^{i \theta} \alpha & -e^{i \theta} \gamma \\
\end{array}\right].
\end{equation}
This gives an infinite family of transformations since $\theta$ can
take any value between $0$ and $2\pi$, and $\alpha, \gamma$ are
$c$- numbers satisfying the constraint (\ref{esc}).
One can get rid of the overall factor by setting $\alpha = 1 = \gamma$  
in Eq.(\ref{est}). 
However, the states then become unnormalized. To restore normalization
one could simply fix $\alpha = 1/{\sqrt{2}} = \gamma$. With this 
choice, the states in (\ref{est}) reduce to the specific form of 
states lying on the equatorial great circle.  So for the 
sake of generality, we shall refrain from assigning any particular value 
to these parameters.
   
Now, we address the following question: Which other orthogonal
pair of qubit states $\{ |\psi\rangle, |{\overline \psi}\rangle \}$ would
transform under $U$ in a similar manner as $\{|0\rangle, |1\rangle\}$? 
More precisely, we wish to find as to which  
class of qubits would satisfy
\begin{equation}
\label{sts}
U|\psi\rangle \rightarrow \alpha |\psi\rangle + e^{i \theta} \alpha
|{\overline \psi}\rangle,~~
U|{\overline \psi}\rangle \rightarrow \gamma |\psi\rangle - e^{i \theta} 
\gamma |{\overline \psi}\rangle,~~ 
\alpha \alpha^* = \gamma \gamma^* = 1/2.
\end{equation}
For this purpose, we start with an arbitrary qubit state 
$|\psi\rangle$ and its orthogonal 
complement $|{\overline \psi}\rangle$ as a superposition of the CB states
\begin{equation}
|\psi\rangle = a |0\rangle + b |1\rangle,~~
|{\overline \psi}\rangle = b^* |0\rangle - a^* |1\rangle,  
\end{equation}
where the non-zero complex numbers obey the normalization condition
$aa^* + bb^* = 1$. 
Substituting the above states in the first expression of 
Eq.(\ref{sts}) gives
\begin{equation}
\label{t1}
U|\psi\rangle = (\alpha a + e^{i \theta} \alpha b^*) |0\rangle
+(\alpha b - e^{i \theta} \alpha a^*) |1\rangle.
\end{equation}
Assuming that $U$ acts linearly on $|\psi\rangle$, we have
\begin{eqnarray}
\label{t2}
U|\psi\rangle &=& a U|0\rangle + b U|1\rangle \nonumber \\
&=& (\alpha a + \gamma b) |0\rangle
+ (e^{i \theta} \alpha a - e^{i \theta} \gamma b) |1\rangle.
\end{eqnarray}
Equating the coefficients in (\ref{t1}) and (\ref{t2}) 
gives
\begin{equation}
\label{con1}
b =  e^{i \theta}\frac{\alpha}{\gamma} b^*,~~~
a + a^* =  ( e^{-i \theta} + \frac{\gamma}{\alpha}) b = e^{-i \theta} b + 
 e^{i \theta} b^*
\end{equation}
Thus, we can state our main result:\\ 
{\em The general equal superposition 
transformation (\ref{sts}) holds for all qubit pairs   
$\{ |\psi\rangle, |{\overline \psi}\rangle \}$ which satisfy 
the constraint (\ref{con1})}.

It can be explicitly checked that unitarity holds for these states. 
Consider two such distinct states $\{ |\psi_1\rangle, |\psi_2\rangle \}$ 
and their orthogonal complements $\{ |{\overline \psi_1}\rangle, 
|{\overline \psi_2}\rangle \}$ which transform according to 
 (\ref{sts}). Taking the inner product, we have
\begin{eqnarray}
\label{ip}
\langle \psi_1|\psi_2\rangle &\rightarrow& \alpha\alpha^* [
\langle \psi_1|\psi_2\rangle + e^{i \theta}
\langle \psi_1|{\overline \psi_2}\rangle + e^{- i \theta}
\langle  {\overline \psi_1}|\psi_2\rangle +
\langle {\overline \psi_1}| {\overline \psi_2}\rangle ], \nonumber \\
\langle {\overline \psi_1}| {\overline \psi_2}\rangle &\rightarrow& 
\gamma\gamma^* [
\langle \psi_1|\psi_2\rangle - e^{i \theta}
\langle \psi_1|{\overline \psi_2}\rangle - e^{- i \theta}
\langle  {\overline \psi_1}|\psi_2\rangle +
\langle {\overline \psi_1}| {\overline \psi_2}\rangle ], 
\end{eqnarray}
where $\alpha \alpha^* = \gamma\gamma^* = 1/2$. To see that these 
states actually satisfy the inner product relations, it is instructive to 
write the complex state parameters as $a=x+iy, b = u+iv$, where $x,y,u,v$ 
are all real. In this notation, a state from this ensemble
reads as
\begin{equation}
|\psi\rangle =\{ \frac{1}{2}(e^{-i\theta}+ \frac{\gamma}{\alpha}) (u+iv) 
+ iy \} |0\rangle + (u+iv) |1\rangle,
\end{equation}
while its orthogonal would be
\begin{eqnarray}
|{\overline \psi}\rangle &=& (u-iv) |0\rangle - \{ 
\frac{1}{2}(e^{i\theta}+ 
\frac{\gamma^*}{\alpha^*}) (u-iv) -
iy \} |1\rangle \nonumber \\
&=& e^{-i\theta} \frac{\gamma}{\alpha} (u+iv) |0\rangle - 
\{ \frac{1}{2}(e^{-i\theta}+
\frac{\gamma}{\alpha}) (u+iv) -iy \} |1\rangle. 
\end{eqnarray}
The inner product rules are then explicitly given as
\begin{eqnarray}
\label{ipr}
\langle \psi_1|\psi_2\rangle & = &\frac{1}{4} ( 6 + e^{i\theta} 
\frac{\gamma}{\alpha} + e^{-i\theta} \frac{\gamma^*}{\alpha^*} ) 
e^{-i\theta}\frac{\gamma}{\alpha} (u_1 + i v_1)(u_2 + i v_2) + y_1y_2 
\nonumber \\
& + & \frac{i}{2} (e^{-i\theta} + \frac{\gamma}{\alpha})[(u_1+iv_1)y_2 - 
(u_2+iv_2)y_1] = {\langle {\overline \psi_1} | {\overline 
\psi_2}\rangle}^*, \nonumber \\
\langle \psi_1|{\overline \psi_2}\rangle &=& i e^{-i\theta}  
\frac{\gamma}{\alpha}[(u_1+iv_1)y_2 -
(u_2+iv_2)y_1] = - {\langle {\overline \psi_1}|\psi_2\rangle}^*.
\end{eqnarray}
Substituting these in (\ref{ip}), we find that the inner product relations 
are indeed preserved.

Our result provides a very convenient unified framework to deduce 
any desired class of equally superposable quantum states. If 
the CB states obey a particular transformation (out 
of the infinite family (\ref{est})), then in a 
single shot we can obtain the entire ensemble of qubits which would 
satisfy the same transformation. To demonstrate its usefulness,  
we present below, two known examples as special cases of our result.

1. {\em Hadamard ensemble }: Choose $\alpha = \frac{1}{\sqrt{2}},~~ \gamma 
= \frac{1}{\sqrt{2}},
~~\theta = 0$. Then ($U \rightarrow U_H$)
\begin{equation}
\label{hct}
U_H|0\rangle = \frac{1}{\sqrt{2}} [ |0\rangle + |1\rangle ],
~~U_H|1\rangle = \frac{1}{\sqrt{2}} [ |0\rangle - |1\rangle ],
\end{equation}
where
\begin{equation}
\label{hm}
U_H= \frac{1}{\sqrt{2}}
\left[\begin{array}{crr}
1 & 1 \\
1 & -1 \\
\end{array}\right].
\end{equation}
This is the well known Hadamard gate with its corresponding 
transformation.
Notice that ${U_H}^2 = I$ 
since $U_H = [\sigma_x + \sigma_z]/{\sqrt{2}}$ 
where $\sigma_x, \sigma_z$ are Pauli matrices. However, in general 
$U^2 \ne I$.

Further, substituting the above choice of the parameters, 
the constraint (\ref{con1}) gives
$b= b^*$, i.e., $b$ is real, and
$a + a^* = 2b$, i.e., $Re(a) = b$.
In terms of the real parameters $x, y, u, v$,  
the above deductions yield $v=0$ and $u=x$. 
Therefore, the qubit states become restricted to   
\begin{equation}
\label{he}
|\psi\rangle = (x+iy)|0\rangle + x|1\rangle,~~
|{\overline \psi}\rangle = x|0\rangle - (x-iy)|1\rangle,~~2x^2+y^2=1.
\end{equation}             
Hence, we have obtained a special class of states which 
transform under the action of the Hadamard matrix $U_H$ via the 
transformation
\begin{equation}
\label{hts}
U_H|\psi\rangle = \frac{1}{\sqrt{2}} [ |\psi\rangle +
|{\overline \psi}\rangle ],~~
U_H|{\overline \psi}\rangle = \frac{1}{\sqrt{2}} [ |\psi\rangle -
|{\overline \psi}\rangle ].
\end{equation}
In other words, this proves the existence of the Hadamard gate (\ref{hm}) 
 for any qubit chosen from the ensemble (\ref{he}).

2. {\em Invariant ensemble }: Choose $\alpha = \frac{1}{\sqrt{2}},~~ 
\gamma = \frac{i}{\sqrt{2}},
~~\theta = \frac{\pi}{2}$. Then ($U \rightarrow U_I$)\\
\begin{equation}
\label{itc}
U_I|0\rangle = \frac{1}{\sqrt{2}} [ |0\rangle + i|1\rangle ],~~
U_I|1\rangle = \frac{1}{\sqrt{2}} [ i |0\rangle + |1\rangle ],
\end{equation}
where
\begin{equation}
\label{sm}
U_I= \frac{1}{\sqrt{2}}
\left[\begin{array}{crr}
1 & i \\
i & 1 \\
\end{array}\right].
\end{equation}
An interesting property of this transformation is that it goes into 
itself, i.e., $U_I|0\rangle \leftrightarrow U_I|1\rangle$ under the 
interchange 
$|0\rangle \leftrightarrow |1\rangle$. For this reason we shall refer to 
it as being `invariant'. The matrix 
$U_I$ is symmetric but not hermitian and 
$U_I^2 = i\sigma_x$ (i.e., the NOT gate) since 
$U_I = [I + i \sigma_x]/{\sqrt{2}}$.

Now, in order to find as to which qubit states would satisfy
\begin{equation}
\label{ste}
U_I|\psi\rangle = \frac{1}{\sqrt{2}} [ |\psi\rangle +
i |{\overline \psi}\rangle ],~~
U_I|{\overline \psi}\rangle = \frac{1}{\sqrt{2}} [i |\psi\rangle +
|{\overline \psi}\rangle ]
\end{equation}
we substitute the above values of $\alpha, \gamma,$ and 
$\theta$ in (\ref{con1}). This yields
$b = b^*$, i.e., $b$ is real, and
$a + a^* = 0$, i.e., $Re(a) = 0$, implying that $a$ is purely imaginary.
Again assuming $a = x+iy$ and $b = u+iv$, 
these constraints give $v=0$ and $x=0$.
Therefore, the qubit states become restricted to
\begin{equation}
\label{ie}
|\psi\rangle = iy|0\rangle + u|1\rangle,~~
|{\overline \psi}\rangle = u|0\rangle + iy|1\rangle,~~~y^2 + u^2 = 1.         
\end{equation}
The above two ensembles were obtained in \cite{mp}
by treating each one separately. 
Here we have shown that they can be deduced from a single general 
ensemble of equally superposed qubits.

The family of transformations which remain invaraint 
under the interchange of $|0\rangle$ and $|1\rangle$ is a subset of the 
general family (\ref{est}), and every member is essentially of 
the type (\ref{itc}). To see this let us 
consider the general transformation (\ref{est}). For this to be invariant
we must have 
$\alpha = - e^{i\theta}\gamma$ and $\gamma = e^{i\theta}\alpha$ which 
implies that $\theta = \pi/2, 3\pi/2$. Substituting 
$\gamma = \pm i\alpha$ in (\ref{est}) we obtain the general form of the
invariant transformation ($U \rightarrow {U_I}^{\prime}$)
\begin{equation}
\label{if}
{U_I}^{\prime}|0\rangle = \alpha [ |0\rangle \pm i|1\rangle],~~~
{U_I}^{\prime}|1\rangle = \alpha [ \pm i |0\rangle + 
|1\rangle],~~~\alpha\alpha^* = 1/2,
\end{equation}
where
\begin{equation}
{U_I}^{\prime}= \alpha
\left[\begin{array}{crr}
1 & \pm i \\
\pm i & 1 \\
\end{array}\right].
\end{equation}
Since $\alpha$ is an overall phase factor, 
it can be readily verified that every member of (\ref{if}) would
lead to exactly the same ensemble (\ref{ie}). Thus, (\ref{itc}) 
can be regarded as a representative of the 
invariant family (\ref{if}).
In what follows, we shall establish our main
result in the context of two other physical principles, namely; the 
no-superluminal signalling condition and the non-increase of entanglement 
under LOCC.    

\section{No-superluminal signalling}

Let us consider the CB states transforming via Eq.(\ref{est}), and 
a qubit state $|\psi\rangle$ transforming under the 
same unitary matrix $U$ via the first expression in (\ref{sts}). We 
first show that if 
$|\psi\rangle$ is completely arbitrary, then this would 
imply superluminal signalling. For this purpose, 
assume that Alice possesses a $3d$ qutrit while Bob has a $2d$ 
qubit and both share the following entangled state:
\begin{equation}
\label{ent1}
|\phi\rangle_{AB} = \frac{1}{\sqrt{3}} \left(|0\rangle_A |0\rangle_B +
|1\rangle_A |\psi\rangle_B +
|2\rangle_A |1\rangle_B  \right).
\end{equation} 
The density matrix of the combined system is defined as $\rho_{AB} =
|\phi\rangle_{AB}
\langle \phi|$. Alice's reduced density matrix can be obtained
by tracing out Bob's part
\begin{eqnarray}
\label{ra1}
\rho_{A} = tr_B(\rho_{AB})& = & \frac{1}{3}~[~ |0\rangle \langle 0| +
|1\rangle \langle 1| + |2\rangle \langle 2| \nonumber \\
&+& a |1\rangle \langle 0| +
a^* |0\rangle \langle 1| +
 b |1\rangle \langle 2| + b^* |2\rangle \langle 1| ~].
\end{eqnarray}
Now Bob applies the above mentioned unitary transformation on his 
qubit states $\{ |0\rangle, |1\rangle, |\psi\rangle \}$ in
Eq.(\ref{ent1}). But,
he does not communicate any information to Alice regarding his operation.  
The shared state then changes to
\begin{equation}
\label{enth}
(I \otimes U) |\phi\rangle_{AB} = |\phi'\rangle_{AB}
= \frac{1}{\sqrt{3}} [~\alpha ~|0 0\rangle + e^{i \theta} \alpha
~|0 1\rangle + \alpha
~|1 \psi\rangle + e^{i \theta} \alpha ~|1 {\overline \psi}\rangle
+ \gamma ~|2 0\rangle - e^{i \theta} \gamma ~|2 1\rangle ~].
\end{equation}
After this operation, Alice's new reduced density matrix becomes
\begin{eqnarray}
\label{ra2}
\rho'_A 
&=& \frac{1}{3}~ [~ |0\rangle \langle 0|+ |1\rangle \langle 1|+
|2\rangle \langle 2| \nonumber \\
&+& \frac{1}{2} (a + e^{-i \theta} b + e^{i \theta} b^* - a^*) |1\rangle 
\langle 0| + \frac{1}{2}
  (a^* + e^{i \theta} b^* + e^{-i \theta} b - a) |0\rangle \langle 1|  
\nonumber \\
&+& \alpha \gamma^* (a - e^{-i \theta} b + e^{i \theta} b^* + a^*) 
|1\rangle \langle 2| +
\alpha^* \gamma (a^* - e^{i \theta} b^*  + e^{-i \theta} b + a) |2\rangle 
\langle 1|~].
\end{eqnarray}
Comparing the coefficients of each term in (\ref{ra1}) and (\ref{ra2}),
it is evident that $\rho'_A \ne \rho_A$ for arbitrary choices of the
parameters $a$ and $b$.
So, in principle, Alice can distinguish between $\rho_A$ and
$\rho'_A$, although Bob has not revealed anything to her about his 
operation. This implies that, with the help of entanglement, superluminal 
communication  has taken place.
But faster-than-light
communication is forbidden by special theory of relativity. Hence, we 
conclude that
the equal superposition transformation does not exist for an arbitrary 
qubit.

If, however, we impose that the no-signalling constraint should not be 
violated, then
$\rho_A$ and $\rho'_A$ should be equal because the action of $U$ is a
trace preserving local operation performed only at Bob's side. 
Comparing coefficients of the term $|1\rangle \langle 0|$ 
in (\ref{ra1}) and (\ref{ra2}) we recover the condition
$a + a^* = e^{-i \theta} b + e^{i \theta} b^* $. From $|1\rangle \langle 
2|$ we have $\alpha \gamma^* (a - e^{-i \theta} b + e^{i \theta} b^* + a^*)
= b $ which yields $2 \alpha \gamma^* e^{i \theta} b^* = b $. 
Substituting $\gamma^* = \frac{1}{2\gamma}$, we get the other constraint 
$b = e^{i \theta}\frac{ \alpha}{\gamma} b^*$. Thus, the no-signalling 
condition gives exactly the same class of states that was obtained 
initially from linearity.

\section{Non-increase of entanglement under LOCC}

Here we shall first show the non-existence of the unitary operation
(\ref{sts}) for an arbitrary $|\psi\rangle$
by considering the fact that local
operations and classical communication cannot increase the entanglement
content of a quantum system. It turns out that, 
$\rho_A$ and $\rho'_A$ above, have equal eigenvalues $(0, 1/3, 2/3)$.
This means that there is no change in entanglement before and 
after the unitary operation. So we consider a different shared 
resource which has been used in \cite{kar, preeti} for studying flipping 
and Hadamard operations,
\begin{equation}
\label{sep1}
{|\Phi\rangle}_{AB} = \frac{1}{\sqrt{1 + b^*b}} [~|0\rangle_A
\frac{{|0\rangle}_{B1}
{|1\rangle}_{B2} - {|1\rangle}_{B1} {|0\rangle}_{B2}}{\sqrt{2}}
+ {|1\rangle}_A \frac{{|0\rangle}_{B1}
{|\psi\rangle}_{B2} - {|\psi\rangle}_{B1} {|0\rangle}_{B2}}{\sqrt{2}}~],
\end{equation}
where the first qubit is with Alice while the other two are at Bob's side.
Repeating the protocol, we obtain Alice's reduced density
operator as
\begin{equation}
\rho_A = \frac{1}{1+b^*b} [~ |0\rangle \langle 0| +
b^*b |1\rangle \langle 1| + b |1\rangle \langle 0|
+ b^* |0\rangle \langle 1| ~].
\end{equation}
The amount of entanglement given by the von Neumann entropy is zero since 
the eigenvalues of $\rho_A$ are $0$ and $1$. This means that the resource 
state (\ref{sep1}) is a product state in the A:B cut.
Now Bob applies the trace preserving general transformation on the last 
particle $(B2)$ in Eq.(\ref{sep1}), which results in 
the state
\begin{eqnarray}
\label{enh}
|\Phi'\rangle_{AB} = \frac{1}{\sqrt{2N}} &[&\gamma |000\rangle - e^{i 
\theta} \gamma |001\rangle - \alpha
|010\rangle - e^{i \theta} \alpha |011\rangle \nonumber \\
&+& \alpha |10\psi\rangle + e^{i \theta} \alpha |10{\overline \psi}\rangle 
- \alpha |1\psi0\rangle - e^{i \theta} \alpha
|1\psi1\rangle ~],
\end{eqnarray}
where $N = 2 + \frac{1}{4} \{(a-a^*)^2 - (e^{-i \theta} b + e^{i \theta} 
b^*) (a+a^*) \}$.
Since $a$ and $b$ are arbitrary, so in general, the above state is
entangled in the A:B cut. This implies
that entanglement has been created by local operation.
However, we know that entanglement cannot be increased by local operations
even if classical communication is allowed.
Therefore, the above contadiction leads us to the conclusion that the 
unitary operator (\ref{esm}) cannot 
perform the same task for an arbitrary qubit, as it does for the 
CB states $|0\rangle$ and $|1\rangle$.

We now derive the conditions under which
the entanglement in the state would remain zero even after the
application of $U$. For this purpose we have to compare
the eigenvalues
of the respective density matrices on Alice's side. So after Bob's
operation
\begin{equation}
\rho'_A = \frac{1}{N} [~|0\rangle \langle 0| +
 (N - 1) |1\rangle \langle 1| + D |1\rangle \langle 0| + D^* |0\rangle 
\langle 1|~],
\end{equation}
where $D = \frac{1}{2} \{ \alpha \gamma^* (a + a^* - e^{-i \theta} b +
e^{i \theta} b^*) + b \}$.
The eigenvalue equation of the above matrix gives two roots, namely,
\begin{equation}
\label{ro}
\lambda_{\pm} = \frac{1}{2} \pm \frac{\sqrt{N^2 - 4 (N - 1 - D D^*)}}{2N},
\end{equation}
In order to maintain the same amount of entanglement in the system
before and after the unitary operation, we should equate these two roots 
$\lambda_{\pm}$ of $\rho'_A$ to the eigenvalues $0$ and $1$ of $\rho_A$.
This furnishes the constraint $D D^* = N - 1 $. Substituting the 
expressions
for $N, D$ and $D^*$, and rearranging the terms, this condition acquires 
the form
\begin{eqnarray}
\label{ddn}
&&[(a+a^*) - (e^{-i \theta}b + e^{i \theta}b^*)]
[ \frac{1}{4}(a+a^*) + \frac{1}{4} (e^{-i \theta}b + e^{i \theta}b^*)
+ \gamma \alpha^* b + \alpha \gamma^* b^* ] \nonumber \\
&&+ 2 (e^{-i \theta} \gamma \alpha^* b^2 + e^{i \theta} \alpha 
\gamma^* b^{*2} + bb^*) = 4 + (a-a^*)^2 - (e^{-i\theta} b + e^{i\theta} 
b^*)(a+a^*).
\end{eqnarray}
Using $\alpha^* = \frac{1}{2\alpha}$, $\gamma^* = \frac{1}{2\gamma}$
on the L.H.S. and adding and subtracting $2aa^*$ on the R.H.S.,
the above relation is recast as
\begin{eqnarray}
\label{ddn1}
&& [(a+a^*) - (e^{-i \theta} b + e^{i \theta} b^*)]
 [ \frac{1}{4}(a+a^*) + \frac{1}{4} (e^{-i\theta} b + e^{i\theta} b^*) +  
\frac{1}{2} (\frac{\gamma}{\alpha} b + 
\frac{\alpha}{\gamma} b^* ) ] \nonumber \\
&& + [ e^{{-i \theta}/2} \sqrt{\frac{\gamma}{\alpha}} b + e^{{i \theta}/2}
\sqrt{\frac{\alpha}{\gamma}} b^* ]^2 = 4 bb^* + [(a+a^*) - (e^{-i 
\theta} b + e^{i \theta} b^*)](a+a^*)
\end{eqnarray}
which can be written more compactly as
\begin{eqnarray}
\label{ddn2}
&& [(a+a^*) - (e^{-i \theta} b + e^{i \theta} b^*)]
 [ - \frac{3}{4}(a+a^*) + \frac{1}{4} (e^{-i\theta} b + e^{i\theta} b^*) +
\frac{1}{2} (\frac{\gamma}{\alpha} b +
\frac{\alpha}{\gamma} b^* ) ] \nonumber \\
&& = - [ e^{{-i \theta}/2} \sqrt{\frac{\gamma}{\alpha}} b - e^{{i 
\theta}/2} \sqrt{\frac{\alpha}{\gamma}} b^* ]^2.
\end{eqnarray}
For convenience let us denote the two terms on L.H.S. by $A$ and $B$. Then 
\begin{equation}
\label{ddn3}
[A] [B] = - [ e^{{-i \theta}/2} \sqrt{\frac{\gamma}{\alpha}} b - e^{{i 
\theta}/2} \sqrt{\frac{\alpha}{\gamma}} b^* ]^2.
\end{equation}
In the above, R.H.S. is either a real positive definite quantity or zero. 
For the L.H.S. to be positive, there, however, exist two possibilities:\\
(i) $ A > 0, B > 0$: If we suppose that both terms in $A$ are 
positive (they are already real), then $a+a^* = (e^{-i\theta} b + 
e^{i\theta} b^*) + C$, where $C$ is a real positive constant. Thus 
$B > 0$ if $ (\frac{\gamma}{\alpha} b + \frac{\alpha}{\gamma} b^* ) > 
 (e^{-i \theta} b + e^{i \theta} b^*) + \frac{3}{2}C$, which certainly is 
possible.
 Similarly if we suppose that both terms in $A$ are negative, then $A > 0$ 
implies
$|e^{-i \theta} b + e^{i \theta} b^*| = |a+a^*| + C$. Thus $B > 0$ if 
$ (\frac{\gamma}{\alpha} b + \frac{\alpha}{\gamma} b^* ) + |a+a^*| > 
\frac{C}{2}$.\\
(ii) $A < 0, B < 0$: In a similar manner, restrictions can be obtained for 
this case.\\
When R.H.S. is identically zero, then Eq.(\ref{ddn3}) would be satisfied 
uniquely if $A = 0, B = 0$. This gives 
$a+a^* = e^{-i \theta} b + e^{i \theta} b^*$ and $b = e^{i\theta} 
\frac{\alpha}{\gamma} b^*$, which are exactly the constraints that 
we have earlier obtained by linearity and no-signalling.
It can be easily checked that the other cases $A=0, B\ne 0$ and 
$A \ne 0, B=0$ cannot exist due to the constraint fixed by R.H.S.
being zero.

The above analysis demonstrates the possibility of existence of more 
solutions
from the principle of non-increase of entanglement under LOCC.
In the case of Hadamard operation, we had obtained a unique solution 
\cite{preeti} from linearity, no-signalling and non-increase of 
entanglement under LOCC.
Here we get a larger set of states with zero entanglement, from 
the last method.
However, we must remember that we are looking for orthogonal 
pairs of states $\{ |\psi\rangle, |{\overline \psi}\rangle \}$
which transform under the unitary operation defined by (\ref{sts}).
In the above, we have considered only $|\psi\rangle$.
Therefore, we must now carry out a similar analysis
with $|{\overline \psi}\rangle$. More precisely, we take the set 
of qubit states $\{ |0\rangle, |1\rangle, |{\overline \psi}\rangle \}$ 
and the shared state as
\begin{equation}
\label{sep2}
{|\Psi\rangle}_{AB} = \frac{1}{\sqrt{1 + a^*a}} [~|0\rangle_A
\frac{{|0\rangle}_{B1}
{|1\rangle}_{B2} - {|1\rangle}_{B1} {|0\rangle}_{B2}}{\sqrt{2}}
+ {|1\rangle}_A \frac{{|0\rangle}_{B1}
{|{\overline \psi}\rangle}_{B2} - {|{\overline \psi}\rangle}_{B1} 
{|0\rangle}_{B2}}{\sqrt{2}}~].
\end{equation}
Then Alice's reduced matrix reads as
\begin{equation}
\rho_A = \frac{1}{1+a^*a} [~ |0\rangle \langle 0| +
a^*a |1\rangle \langle 1| - a^* |1\rangle \langle 0|
- a |0\rangle \langle 1| ~].
\end{equation}
Bob now applies $U$ on the states
$\{ |0\rangle, |1\rangle, |{\overline \psi}\rangle \}$
 of his last qubit, thereby changing
the shared state to 
\begin{eqnarray}
\label{enhb}
|\Psi'\rangle_{AB} = \frac{1}{\sqrt{2{\cal N}}} &[&\gamma |000\rangle - 
e^{i \theta} \gamma |001\rangle - \alpha
|010\rangle - e^{i \theta} \alpha |011\rangle \nonumber \\
&+& \gamma |10\psi\rangle - e^{i \theta} \gamma |10{\overline \psi}\rangle
- \alpha |1{\overline \psi}0\rangle - e^{i \theta} \alpha
|1{\overline \psi}1\rangle ~],
\end{eqnarray}
where ${\cal N} = 2 - \frac{1}{2} \{\gamma \alpha^*b(a + a^* + e^{-i 
\theta} b - e^{i \theta} b^*) + \alpha \gamma^* b^* (a + a^* - 
e^{-i \theta} b + e^{i \theta} b^*) \}$.
The corresponding reduced density matrix at Alice's end becomes
\begin{equation}
\rho'_A = \frac{1}{\cal N} [~|0\rangle \langle 0| +
 ({\cal N} - 1) |1\rangle \langle 1| + {\cal D}
|1\rangle \langle 0| + {\cal D}^* |0\rangle \langle 1|~],
\end{equation}
where ${\cal D} = \{ \frac{1}{4}(a - a^* - e^{-i \theta} b - e^{i \theta}
b^*) - \frac{a^*}{2} \}$.
Like the previous case, this matrix has the following two eigenvalues  
\begin{equation}
\label{ro2}
\lambda_{\pm} = \frac{1}{2} \pm \frac{\sqrt{{\cal N}^2 - 4 ({\cal N} - 1 - 
{\cal D} {\cal D}^*)}}{2{\cal N}}.
\end{equation}
Equating $\lambda_{\pm}$ to the eigenvalues $0$ and $1$ of $\rho_A$ gives 
the constraint
${\cal D} {\cal D}^* = {\cal N} - 1$ which can be expanded as
\begin{eqnarray}
\label{ddnb}
&& [(a+a^*) - (e^{-i \theta} b + e^{i \theta} b^*)]
 [ - \frac{3}{8}(a+a^*) - \frac{1}{8} (e^{-i\theta} b + e^{i\theta} b^*) +
\frac{1}{2} (\frac{\gamma}{\alpha} b +
\frac{\alpha}{\gamma} b^* ) ] \nonumber \\
&& = - [ e^{{-i \theta}/2} \sqrt{\frac{\gamma}{\alpha}} b - e^{{i
\theta}/2} \sqrt{\frac{\alpha}{\gamma}} b^* ]^2.
\end{eqnarray} 
Interestingly, this is a new restriction on the expression on R.H.S.
This has to be consistent with the earlier restriction (\ref{ddn2}). 
Therefore equating (\ref{ddnb}) with (\ref{ddn2}) 
renders $a+a^* = e^{-i \theta} b + e^{i \theta} b^*$. Substituting this 
in either (\ref{ddn2}) or (\ref{ddnb}) yields $b = e^{i\theta}
\frac{\alpha}{\gamma} b^*$. Thus we finally obtain a unique solution which 
is exactly the constraint (\ref{con1}) that defines our equal 
superposition ensemble.

We remark that such a situation was not encountered in the case of the
Hadamard operation \cite{preeti}. The reason is that the Hadamard 
transformation on $|{\overline \psi}\rangle$ is not independent since it 
can be obtained from the
Hadamard transformation on the states $\{ |0\rangle, |1\rangle, 
|\psi\rangle \}$
by using the special property of the Hadamard operator, namely, 
${U_H}^2 = I$.
However, in the present scenario (and in general), 
$U|{\overline \psi}\rangle$ cannot be deduced from 
$\{ U|0\rangle, U|1\rangle, U|{\overline \psi}\rangle \}$.
So it is necessary to take $U|{\overline \psi}\rangle$  
into consideration, 
although whether this would provide
some new restriction or not depends on the particular situation.
For example, if we proceed with $|{\overline \psi}\rangle$, then linearity
and no-signalling give nothing new but the same constraint 
(\ref{con1}) which was obtained from $|\psi\rangle$. 
However, in the framework of non-increase of entanglement under LOCC,
this indeed yields a different condition (\ref{ddnb}), thereby forcing the
set of solutions to a single unique solution.
In view of the above, we are now in a position to make a stronger 
statement regarding our main result:
 
{\em Any pair of qubit states $\{ |\psi\rangle, 
|{\overline \psi}\rangle \}$ can be equally superposed via the unitary 
operation (\ref{sts}) if and only if they satisfy the constraint 
(\ref{con1}) }.

\section{Quantum Random Walks} 

In the previous sections, we have obtained by different methods the class 
of qubit states 
which transform under the action of the unitary matrix $U$ in a manner 
similar to Eq. (\ref{est}). As an application of this transformation 
(\ref{est}), we are now going to study the quantum random 
walk associated with it. A particularly nice detailed survey of 
quantum walks has been 
given by Kempe \cite{kempe}, while \cite{amba} is a short review devoted 
to their applications to algorithms. The Hadamard
matrix $U_H$ has been widely 
used as a balanced coin (translation to the left or to the 
right with equal probability) to 
study the properties of a discrete-time quantum random walk 
(QRW)\cite{nayak}.
For example, the probability of finding the particle at a particular site
after $T$ steps of the walk have been investigated in detail. 
The Hadamard coin gives an asymmetric probability distribution for the
QRW on a $1d$ line. This is because the Hadamard coin 
treats the two CB states 
differently; it multiplies the phase by $-1$ only in the case of 
$|1\rangle$. It has also been pointed out \cite{kempe, amba} that
if the Hadamard coin is replaced with the more symmetric coin
$U_I$, then the probability distribution becomes symmetric. However, our 
analysis shows that this is not the case, even though $U_I$  
treats both $|0\rangle$ and $|1\rangle$ in a symmetrical way.
This also motivates us to investigate the discrete-time QRW from a more 
general point of view. We shall study the behaviour of the walk by taking the 
general unitary matrix $U$ given by Eq.(\ref{esm}) as our balanced coin. 
Subsequently, we shall 
comment on some interesting features that these walks share. 

Consider a particle localized at position $z$ on a $1d$ line. 
The Hilbert space ${\cal H}_P$ is 
spanned by basis states $|z\rangle$, where $z$ is an integer. This 
position Hilbert space is augmented by a coin space ${\cal H}_C$ spanned 
by the
two CB states $|0\rangle$ and $|1\rangle$. To avoid confusion with the 
position states, we now introduce a change of notation, and instead denote 
the CB states as $|\uparrow\rangle$ and $|\downarrow\rangle$.
The total state of the particle lies in the Hilbert space 
${\cal H} = {\cal H}_C \otimes {\cal H}_P$. 

The first step of the random walk is a 
rotation in the coin space. We follow a procedure similar to what was 
adopted for the Hadamard walk \cite{kempe}.
In our general scenario, the matrix $U$ given by (\ref{esm}) serves as 
the coin, with the following action (cf. Eq.(\ref{est}))
\begin{equation}
\label{ct}
U|\uparrow\rangle = \alpha |\uparrow\rangle + e^{i\theta} \alpha 
|\downarrow\rangle ,~~~  
U|\downarrow\rangle = \gamma |\uparrow\rangle - e^{i\theta} \gamma
|\downarrow\rangle.
\end{equation}
The rotation is followed by translation with the application of the 
unitary operator
\begin{equation}
\label{cso}
S = \ua \langle\uparrow| \otimes \sum_z |z+1\rangle \langle z|
+  \da \langle\downarrow| \otimes \sum_z |z-1\rangle \langle z|
\end{equation}
in the position space ${\cal H}_P$. Note that $S$ is a `conditional' 
translation operator since it moves the particle by 
one unit to the right if the coin state is $\ua$, and to the left 
if it is $\da$
\begin{equation}
\label{cot}
S \ua \otimes |z\rangle = \ua \otimes |z+1 \rangle,
 ~~~~S \da \otimes |z\rangle = \da \otimes |z-1 \rangle.
\end{equation}
The particle is subjected to these two alternating 
unitary transformations.
Therefore, the QRW of $T$ steps is defined as the transformation
${\cal A}^T$, where $\cal A$ acts on the total Hilbert space ${\cal H}$ 
and is given 
by
\begin{equation}
\label{cut}
{\cal A} = S(U\otimes I)
\end{equation}
To start with, let the particle be in the 
$|\uparrow\rangle$ coin 
state and located at the position $0$. Thus the total initial state is 
denoted by $|\phi\rangle = \ua \otimes |0\rangle$.
Let us now evolve the walk, for a few steps, under successive action of 
the operator $\cal A$:
\begin{eqnarray}
\label{evo}
|\phi\rangle &\rightarrow&  \alpha \ua \otimes |1\rangle + 
e^{i\theta} \alpha\da \otimes |-1\rangle \nonumber \\
&\rightarrow& \alpha^2 \ua \otimes |2\rangle + e^{i\theta}(\alpha^2 \da + 
\alpha \gamma \ua) \otimes |0\rangle - e^{2i\theta} \alpha\gamma 
 \da \otimes |-2\rangle \nonumber \\
&\rightarrow& \alpha^3 \ua\otimes |3\rangle +  e^{i\theta}(\alpha^3 \da
+ 2 \alpha^2 \gamma \ua) \otimes |1\rangle - e^{2i\theta} \alpha \gamma^2 
\ua \otimes |-1\rangle \nonumber \\ 
&+& e^{3i\theta} \alpha \gamma^2 \da 
\otimes |-3\rangle \nonumber \\
&\rightarrow& \alpha^4 \ua \otimes |4\rangle +  e^{i\theta}(\alpha^4 \da +
 3\alpha^3 \gamma \ua)\otimes |2\rangle + e^{2i\theta}(\alpha^3 \gamma 
 \da - \alpha^2 \gamma^2  \ua) \otimes |0\rangle \nonumber \\
&-& e^{3i\theta} (\alpha^2 \gamma^2  \da - \alpha \gamma^3 \ua) 
\otimes |-2\rangle - e^{4i\theta}\alpha \gamma^3 \da \otimes 
|-4\rangle
\end{eqnarray}
After $T$ iterations, the particle is in an entangled state, say 
$|{\phi}_T\rangle$.
The probability of finding the particle at a particular site $z$ is  
given by
\begin{equation}
P_z = |(\langle \uparrow| \otimes \langle z|)|{\phi}_T\rangle|^2 
+ |(\langle \downarrow| \otimes \langle z|)|{\phi}_T\rangle|^2.
\end{equation}
Let us analyze, step by step, the spatial probability distribution 
of the walk.

After $T=1$: If we measure the position space after the first step,
then the particle can be found at the site $1$ with probability
$\alpha\alpha^*$ and at the site $-1$ with the same probability.
Since we already know that 
$\alpha\alpha^* = 1/2$ (normalization), so the particle moves 
with equal probability, one step to the right and one to the left of its 
original position.
The walk is therefore, unbaised, just like the usual Hadamard walk.

After $T=2$: The probabilities of finding the particle at positions
$2$, $-2$ and $0$ are respectively,
\begin{equation}
P_2 = \alpha^2 \alpha^{2*} = (\alpha \alpha^*)^2 = 1/4, ~~~
P_{-2} = \alpha \alpha^* \gamma \gamma^* = 1/4,~~~
P_0 = \alpha^2 \alpha^{2*} + \alpha \alpha^* \gamma \gamma^* = 1/2.
\end{equation} 
This step is also similar to the case of classical walk since the $P's$ 
are symmetrically distributed.

After $T=3$: The distribution is
\begin{eqnarray}
\label{p3}
P_3 &=& \alpha^3 \alpha^{3*} = (\alpha\alpha^*)^3 = 
1/8,~~~P_{-3} = \alpha \alpha^* 
\gamma^2 \gamma^{2*} = 1/8, \nonumber\\
P_1 &=& (\alpha\alpha^*)^3 + 4 (\alpha\alpha^*)^2 \gamma\gamma^* = 
5/8,~~~
P_{-1} = \alpha\alpha^* (\gamma\gamma^*)^2 = 1/8.
\end{eqnarray}
After the third step, the quantum walk begins to deviate from its 
classical counterpart. Although $P_3 = P_{-3}$, note that
$P_1 \ne P_{-1}$. So the walk starts to be asymmetric, 
drifting towards the right since the site $1$ has greater 
probalility. 

After $T=4$: Similarly, upon measuring the position space after four 
iterations, we get the following asymmetric distribution
\begin{equation}
P_4 = 1/16, ~P_{-4} = 1/16, ~P_2 = 5/8, ~P_{-2} = 1/8, ~P_0 = 1/8.
\end{equation}
Again, this differs from the symmetric classical probability distribution 
$P_4 = 1/16, ~P_{-4} = 1/16, ~P_2 = 1/4, ~P_{-2} = 1/4, ~P_0 = 3/8$.
Proceeding in a similar way one can check the veracity of the foregoing
conclusions by considering more steps of iterations. 
Clearly, the parameter $\theta$ which appears in the phase factor
does not contribute to the probabilities. Also since 
$\alpha\alpha^* = \gamma\gamma^*$, so for the purpose of probability 
distribution, only one of the parameters may be regarded as independent.

It is observed that the spatial probability distribution of the QRW 
corresponding to 
the general matrix $U$ is asymmetrical and coincides exactly with that of 
the already known Hadamard walk. This means that every unitary 
transformation in 
which the qubit CB states are equally weighted,
leads to the same probability distribution if the particle is taken 
in the same initial state. Therefore, we infer that even the symmetric 
coin $U_I$ induces an asymmetrical walk.
In fact, it can be argued easily as to why a  
symmmetric probability distribution for the initial state 
$\ua \otimes |0\rangle$ (or $\da \otimes |0\rangle$) is impossible.
Let us refer to the distribution (\ref{p3}) after three iterations. If we 
want to make it symmetric, we must have $P_1 = P_{-1}$. This implies that 
\begin{equation}
\alpha\alpha^* [(\alpha\alpha^*)^2 + 4 \alpha\alpha^*
\gamma\gamma^* - (\gamma\gamma^*)^2] = 0.
\end{equation}
This equation cannot be satisfied since we know that $\alpha \alpha^* = 
1/2$, and the term in the bracket equals $1$.
So L.H.S. can never be zero. \\

The direction of drift in the walk depends on the initial coin state and
the bias is the result of quantum interference. So phases play a very 
crucial role in inducing asymmetry. This bias can, however, be 
taken care of if we again allow interference, so that the effect of the 
earlier superposition is negated.
Thus, in order to make the walk symmetric or unbiased, we must take a 
superposition of $\ua$ and $\da$ as the initial coin state. However, we shall
not assume apriori that $\ua$ and $\da$ are superimposed with equal 
probability.
For our general approach, we shall rather superimpose them with arbitrary 
amplitudes and obtain restrictions under which we can get a symmetric 
distribution. 
So we start the walk in the state
\begin{equation}
|\phi^{\prime}\rangle = (x \ua + y \da) \otimes |0\rangle, ~~~~x x^* + y 
y^* = 1
\end{equation}
($x$ and $y$ are, in general, complex numbers) and let it evolve under the 
repeated action of the operator $\cal A$, as was done earlier.

After $T=1$, the state becomes 
\begin{eqnarray}
|{\phi}_1\rangle &=& (x \alpha + y \gamma)\ua \otimes |1\rangle + 
e^{i\theta}(x \alpha - y \gamma) 
\da \otimes |-1\rangle \nonumber \\
&=& A \ua \otimes |1\rangle + e^{i\theta} 
B \da\otimes |-1\rangle
\end{eqnarray}
We now demand that the particle should be found at sites $1$ and $-1$ with 
equal probability. This gives the constraint
\begin{equation}
A A^* = B B^* = 1/2
\end{equation}
which can be recast in terms of the transformation parameters as
\begin{equation}
x y^* \alpha \gamma^* + x^* y \alpha^* \gamma = 0.
\end{equation}
Clearly, this holds only if 
$x y^* \alpha \gamma^*$ is purely imaginary.

After $T=2$, the state of the particle becomes
\begin{equation}
|{\phi}_2\rangle = 
\alpha A\ua\otimes|2\rangle + e^{i\theta}(\alpha A\da + \gamma B\ua) 
\otimes |0\rangle - e^{2i\theta}\gamma B\da\otimes|-1\rangle
\end{equation}
and the probabilities are 
\begin{equation}
P_2 = \alpha\alpha^* AA^* = 1/4, ~~~P_{-2} = \gamma\gamma^* BB^* = 1/4,~~~
P_0 = \alpha\alpha^* AA^* + \gamma\gamma^* BB^* = 1/2.
\end{equation}     \\
After $T=3$, the state evolves into
\begin{eqnarray}
|{\phi}_3\rangle
&=&\alpha^2 A \ua\otimes |3\rangle + e^{i\theta} (\alpha^2 
A\da + 2x\alpha^2\gamma\ua)\otimes|1\rangle \nonumber \\
&-& e^{2i\theta}(\gamma^2 B\ua + 
2y\gamma^2\alpha\da)\otimes|-1\rangle + e^{3i\theta}\gamma^2 B 
\da\otimes|-3\rangle
\end{eqnarray}
The probabilities for the odd sites are
\begin{eqnarray}
P_3 &=& (\alpha\alpha^*)^2 AA^* = 1/8, ~~~P_{-3} = (\gamma\gamma^*)^2 
BB^* = 1/8 \nonumber \\
P_1 &=& (\alpha\alpha^*)^2 AA^* + 4 x x^* (\alpha\alpha^*)^2\gamma\gamma^*
= 1/8 + xx^*/2\nonumber \\
P_{-1} &=& (\gamma\gamma^*)^2 BB^* + 
4y y^*(\gamma\gamma^*)^2\alpha\alpha^* = 1/8 + yy^*/2 
\end{eqnarray}
For symmetry, $P_1 = P_{-1}$ which, in turn, implies that
$x x^* = y y^*$. But from
normalization, we have $x x^* + y y^* = 1$. This restricts the value to
$xx^* = yy^* = 1/2$. So the two amplitudes are equal, upto a phase 
factor, leading to an equal superposition of $\ua$ and $\da$.
We have thus found that in order to make the quantum walk
associated with the matrix $U$ symmetric, it is necessary to take 
an `equally' superposed coin state in such a way 
that $xy^*\alpha\gamma^*$ is purely imaginary. We present a new 
example to illustrate this situation.\\

{\em Example:} Choose $\alpha = \gamma = \frac{1+i}{2}$ and 
$\theta = \frac{3\pi}{2}$. The transformation (\ref{ct}) becomes
\begin{equation}
U \ua = \frac{1+i}{2} (\ua - i \da), ~~~U \da = \frac{1+i}{2} (\ua + i 
\da)
\end{equation}
This has the features of a `hybrid' between the Hadamard and the Invariant 
transformations discussed earlier. As expected, this gives an asymmetric 
walk. However, if we take the coin state in a superposition with 
amplitudes $x = \frac{1}{\sqrt{2}}$ and $y = \frac{-i}{\sqrt{2}}$, then 
the condition that $xy^*\alpha\gamma^*$ is purely imaginary is satisfied.
Thus upon evolving the walk with the initial state
$\frac{1}{\sqrt{2}} (\ua - i \da) \otimes |0\rangle$, we do get the 
symmetric probability distribution which coincides with the classical 
one.\\

{\bf Unbalanced coin: }\\
We have seen above that unitary balanced coins
lead to a symmetric walk after appropriate superposition.
Now we shall show that even unbalanced coins can yield an unbiased 
walk under similar restrictions. 

Consider the unequal superposition transformation given in \cite{mp} 
\begin{equation}
\label{ust} 
{\cal U} \ua = p \ua + q \da, ~~~{\cal U} \da = q^* \ua - p^* \da,~~~pp^* 
+ qq^* = 1,
\end{equation}
where, in general, $p p^* \ne q q^*$ and the unitary matrix $\cal U$ is 
given by
\begin{equation}
\label{usm}
{\cal U}= 
\left[\begin{array}{crr}
p & q^* \\
q & -p^* \\
\end{array}\right].
\end{equation}
Following the same procedure, let us start the walk in the superposed 
state
\begin{equation}
|\Phi^{\prime}\rangle = (r \ua + s \da) \otimes |0\rangle, ~~~~r r^* + s
s^* = 1,
\end{equation}
$r, s$ being non-zero $c-$ numbers.
After $T=1$, the state becomes
\begin{eqnarray}
|{\Phi}_1\rangle &=& (r p + s q^*)\ua \otimes |1\rangle +
(r q - s p^*) \da \otimes |-1\rangle \nonumber \\
&=& E \ua \otimes |1\rangle + F \da\otimes |-1\rangle.
\end{eqnarray}
For a symmetric probability distribution, we must have
\begin{equation}
E E^* = F F^* = 1/2,
\end{equation}
which leads to the constraint
\begin{equation}
\label{pqcon}
r s^* p q + r^* s p^* q^* = 0,~~~ r r^* = s s^*.
\end{equation}
This implies that like the previous case, here also the coin state must 
be in {\em  equal} superposition (upto a phase) of $\ua$ and $\da$.

After $T=2$, the particle is in the entangled state 
\begin{equation}
|{\Phi}_2\rangle =
p E \ua \otimes |2\rangle + (q E \da + q^* F \ua)
\otimes |0\rangle - p^* F \da \otimes |-2\rangle
\end{equation}
with the probability distribution
\begin{equation}
\label{pdu2}
P_2 = p p^* E E^* = \frac{1}{2} p p^*,~~~ P_{-2} = p p^* F F^* = 
\frac{1}{2} p p^*,~~~ P_0 = q q^* E E^* + q q^* F F^* = q q^*,
\end{equation} 
After $T=3$, the entangled state is read as
\begin{eqnarray}
|{\Phi}_3\rangle &=&
p^2 E \ua\otimes|3\rangle + p q E \da \otimes |1\rangle +(q q^* E + p q^* 
F) \ua \otimes |1\rangle \nonumber \\
&+& (-p^* q E + q q^* F) \da \otimes|-1\rangle - p^* q^* F 
\ua\otimes|-1\rangle + {p^*}^2 F \da \otimes |-3\rangle,
\end{eqnarray}
and the associated probabilities are
\begin{eqnarray}
P_3 &=& p^2 {p^2}^* E E^* = (p p^*)^2 E E^* = \frac{1}{2} (p p^*)^2 , 
~~~P_{-3} = (p p^*)^2 F F^* = \frac{1}{2} (p p^*)^2, \nonumber \\
P_1 &=& p p^* q q^*  + \frac{1}{2} (q q^*)^2, ~~~P_{-1} = p p^* q q^* + 
\frac{1}{2} (q q^*)^2.
\end{eqnarray}
Clearly, the above distribution is symmetric.
One can continue like this for large times.\\

{\em Example:} Let
$p = \frac{\sqrt{3}}{2}$, $q = \frac{1}{2}$, $r = \frac{1}{\sqrt{2}}$
and $s = \frac{i}{\sqrt{2}}$.
It can be checked that the constraint (\ref{pqcon}) holds for this choice. 
The associated probabilities are  
\begin{eqnarray}
&&T=1 : P_1 = P_{-1} = 1/2, \nonumber \\
&&T=2 : P_2 = P_{-2} = 3/8,~~~ P_0 = 1/4, \nonumber \\
&&T=3 : P_3 = P_{-3} = 9/32,~~~P_1 = P_{-1} = 7/32.
\end{eqnarray}
Hence we get a new symmetric distribution which is different from the 
classical one. This example demonstrates new possiblities in the quantum 
world which have no classical analoques.
The distribution depends only on the values of $p$ and $q$, after the 
initial constraint (\ref{pqcon}) is satisfied. For 
$p = q = \frac{1}{\sqrt{2}}$, we recover the Hadamard walk. So this walk 
can be thought of as a `generalized' Hadamard walk for unequal amplitudes
$p$ and $q$.   

\section{Conclusions}

In this work, we have established that it is not possible to 
create a superposition with equal probabilities, of an arbitrary qubit 
state and its orthogonal.
The class of states for which this can be 
achieved is presented. In addition, by using the 
principles of no-superluminal signalling and non-increase of entanglement 
under LOCC we have shown that this is the only set of qubits which would 
satisfy the equal superposition transformation. In other words, a qubit 
state and its complement can be equally superposed {\em if and only if }
they belong to the aforementioned ensemble.  

The quantum random walk associated with this general unitary equal 
superposition transformation has been investigated from the point of 
view of probability distribution of a particle. We have found that 
the entire family leads to the same asymmetric 
distribution. This implies that even the symmetric transformation 
(\ref{itc}) gives an 
asymmetric walk. It may be mentioned that apart from the CB vectors, any 
state from our ensemble specified by (\ref{con1}),
can be used as a coin state to study the 
evolution of the walk. The measurement on the coin register would then 
have to be carried out in the $\{ |\psi\rangle, |{\overline \psi}\rangle 
\}$ basis.
We have also obtained conditions under which equal and unequal 
superpositions would yield unbiased walks. To illustrate this, a few 
examples have been presented.  
We have analysed the evolution explicitly only upto four iterations. It 
would be interesting to simulate the walk associated with the 
unbalanced coin for large times and study the mixing time and other 
properties.

{\bf Acknowledgement:} I thank G. Kar for useful discussions.

\end{document}